\documentstyle[sprocl]{article}
\def\Journal#1#2#3#4{{#1} {\bf #2}, #3 (#4)}
\def\NPB{{\em Nucl. Phys.} B}
\def\PLB{{\em Phys. Lett.}  B}
\def\PL{\em Phys. Lett.}  
\def\PRD{{\em Phys. Rev.} D}
\newcommand{\beq}{\begin{equation}}
\newcommand{\eeq}{\end{equation}}
\newcommand{\bea}{\begin{eqnarray}}
\newcommand{\eea}{\end{eqnarray}}
\def\Re{\hbox{Re}\,}
\def\Im{\hbox{Im}\,}

\def\de{\delta}
\def\ep{\epsilon}

\def\si{\sigma}

\def\ps{\psi}

\def\Ga{\Gamma}
\def\De{\Delta}

\def\fr#1#2{{{#1} \over {#2}}}

\def\half{{\textstyle{1\over 2}}}
\def\frac#1#2{{\textstyle{{#1}\over {#2}}}}

\def\lsim{\mathrel{\rlap{\lower4pt\hbox{\hskip1pt$\sim$}}
    \raise1pt\hbox{$<$}}}
\def\gsim{\mathrel{\rlap{\lower4pt\hbox{\hskip1pt$\sim$}}
    \raise1pt\hbox{$>$}}}
\def\sqr#1#2{{\vcenter{\vbox{\hrule height.#2pt
         \hbox{\vrule width.#2pt height#1pt \kern#1pt
         \vrule width.#2pt}
         \hrule height.#2pt}}}}

\begin{document}

\begin{flushright}
IUHET 340\\
August 1996
\end{flushright}
\medskip

\title{PROPOSED TESTS OF CPT SYMMETRY USING $D$ MESONS}
\author{DON COLLADAY and V.\ ALAN KOSTELECK\'Y }
\address{Indiana University, Physics Department,
Bloomington, IN 47401, USA}
\maketitle\abstracts{
Parameters describing CPT violation are extracted from a
variety of rate asymmetries in the neutral-$D$ system.
The precision to which these parameters could be measured
in present and planned machines is examined.}

CPT symmetry is known to be preserved by local, relativistic,
point-particle field theories.\cite{sw}
This symmetry has been experimentally verified to a high degree
of accuracy using the neutral-kaon system.\cite{pdg}
The precision attained in such experiments makes CPT 
an excellent probe of fundamental particle physics.
Additional motivation to search for CPT violation 
comes from a proposed mechanism through which string 
theory might lead to violations of CPT 
at a level comparable to recent experimental bounds.\cite{kp1,kp2}

The interferometric nature of the $K$ system 
that makes sensitive CPT tests possible is shared 
by both the $D$ and $B$ systems.\cite{kp2}$^-$\cite{ck2}
In this talk,
some prospects for using the $D$ system 
for CPT tests are briefly described.
Further details and our notational conventions
can be found in ref.\ 6. 

The $D$ system might naively be 
viewed as less useful for CPT studies 
due to the small expected size of the mixing parameter $x$ 
characterizing the decay time relative to the mixing time.
However,
the suppression by $x$ of the complex parameter $\de_D$
describing indirect $D$-system violation
can be an advantage in isolating other quantities,\cite{ck2}
including parameters for direct $D$-system CPT violation 
and the parameter $\de_K$ for indirect $K$-system CPT violation.
The $K$-system dependence arises when 
$D$ decays into neutral kaons are studied.

Two types of experiment are considered:
one involving decays of uncorrelated $D$ mesons,
and the other involving the production of $\ps(3770)$
with subsequent decay into correlated neutral-$D$ pairs.
The variables of experimental interest are asymmetries
in decay rates,
which have been classified.\cite{ck2}

An example of an asymmetry isolating $\de_D$
can be constructed from
the rates of double-semileptonic decays 
of correlated $D$ mesons
to a final state $f$ in one channel 
and its charge conjugate $\overline{f}$ in the other:
\beq
A_{f,\overline f} \equiv \fr
{\Ga^+(f,\overline f) - \Ga^-(f,\overline f)}
{\Ga^+(f,\overline f) + \Ga^-(f,\overline f)}
\simeq x(\Re \de_D + 2 \Im \de_D)
\quad .
\label{jk}
\eeq
The superscripts $+$ and $-$ 
on the decay rates $\Ga$ indicate that the decay 
into $f$ occurs, respectively,
before and after the decay into $\overline{f}$.
Note the suppression of $\de_D$ by the factor $x$.

To obtain an example of an asymmetry 
involving $\de_K$ for kaons,
consider the following combination of 
total decay rates $R_S$ of uncorrelated $D^0$ mesons 
into final states containing $K_S$ 
and $\overline R_S$ of $\overline{D^0}$ mesons 
into the same final states:
\beq
A_S \equiv \fr {\overline R_S - R_S}
{\overline R_S + R_S}
\quad .
\eeq
The difference between this asymmetry
and an analogous one involving $K_L$ final states is 
\bea
A_L - A_S & = &
- 4 \Re \de_K 
+ 2\Re (\overline x_K - x_K)
+ 2 x \Re (\ep_D - \ep_K - y_K)
\nonumber \\
&  & + 4 x \left[\Im (\ep_K + \ep_D) + 
\fr{\Im F_K}{\Re F_K}\right]
\nonumber \\
&  & + 4 x^2 \Re (\de_D + \de_K - \half (\overline{x}_K - x_K))
\nonumber \\
& & - 2 x^2 \Im (\de_D - \de_K + \half 
(\overline{x}_K - x_K))
\quad .
\label{ig}
\eea
The various quantities in this equation parametrize
direct and indirect T and CPT violation in the $D$ and
$K$ systems.
Applying the condition $x \ll 1$ of small $D$-system mixing
suppresses all but the first two terms, 
thereby specifically isolating CPT-violating parameters
in the $K$ system.
The first term, $-4\Re \de_K$, 
parameterizes indirect CPT violation in the $K$ system.
The second term, $2\Re (\overline x_K - x_K)$, 
is a measure of simultaneous direct CPT breaking 
and violation of the $\De C = \De Q$ rule.

Given an expression for an asymmetry
depending on some parameter,
the number of events needed to measure that parameter
to within one standard deviation $\si$ can be estimated
directly if experimental acceptances and backgrounds
are neglected.\cite{rosner}
Consider first Eq.\ \ref{jk}.
The number $N_{\psi (3770)}$ of $\psi (3770)$ 
events required is determined using the inverse 
branching ratios of the $\psi (3770)$
into the relevant double-semileptonic final states
and summing contributions to the asymmetry.
This gives
\beq
N_{\psi (3770)}(\Re\de_D +2 \Im\de_D ) \simeq
\fr{9}{x^2\si^2}
\simeq \fr {3600}{\si^2}
\quad .
\label{npsi}
\eeq
The numerator 3600 assumes $x \simeq 0.05$,
which is close to the experimental limit and 
illustrates the best possible scenario
for bounding $\de_{D}$.
The value of $x$ may theoretically be this large if long-distance
dispersive effects dominate\cite{geo} 
or if certain extensions of the 
standard model are invoked.\cite{blay}

Next,
an estimate is made of the accuracy that can be achieved
in the measurement of the parameters in Eq.~\ref{ig}.
The inverse branching ratio of the $D$ meson 
into the relevant final state,
which is typically of the order of several percent,
is needed as input.
The number $N_D$ of uncorrelated $D$ mesons
required to measure $\Re \de_K$ 
to an accuracy of one standard deviation $\si$ is 
\beq
N_D (\Re \de_K) \simeq 
\fr1{16\si^2 {\rm BR}(D^0 \rightarrow \overline{K^0} + any)}
\quad .
\label{x}
\eeq  
For simplicity,
any violations of the $\De C = \De Q$ rule 
have been assumed to be independent of direct CPT violation.

Prospects for experimentally 
studying CPT symmetry using the neutral-$D$ system 
have been presented.
For a sample of $10^8$ $\psi(3770)$ events,
which could be generated by a $\tau$-charm factory,
a theoretical bound could be placed 
on indirect $D$-system CPT violation 
at about the $10^{-2}$ level if $x$ is large enough.
Furthermore,
the small size of $x$ can be used to advantage in extracting 
the parameter $\de_K$ for indirect CPT violation in the $K$ system.
Its real part could in principle already be bounded 
to the $10^{-2}$ level or better 
using the currently available $10^5$ reconstructed 
neutral-$D$ events
and could be significantly improved in the near future.
This provides a different method of measuring $\de_K$.
We remark that it is also possible to extract certain  
parameters for direct CPT violation in the $D$ system
using similar methods.\cite{ck2}

\end{document}